              \documentstyle[12pt]{article}
\begin{document}\thispagestyle{empty}\begin{flushright}

INR--820/93 \\
OUT--4102--45 \\
                                     \end{flushright}

                                                     \begin{center}{\large\bf
Connections between deep-inelastic and annihilation processes  \\[4pt]
at next-to-next-to-leading order and beyond$^1$                }
                                                      \vfill{\bf
D.~J.~Broadhurst$^2$}
                                                      \vglue 2mm
Physics Department,
Open University,                                      \\
Milton Keynes, MK7 6AA, UK
                                                      \vglue 3mm{\bf
A.~L.~Kataev$^3$}
                                                      \vglue 2mm
Institute for Nuclear Research
of the Academy of Sciences of Russia,                 \\
117312 Moscow, Russia                                 \end{center}\vfill

\newcommand{\rd}{{\rm d}}
\newcommand{\Ca}{C_{\rm A}}
\newcommand{\Cf}{C_{\rm F}}
\newcommand{\Sa}{S_{\rm A}}
\newcommand{\Sf}{S_{\rm F}}
\newcommand{\Nf}{N_{\rm F}}
\newcommand{\Tf}{T_{\rm F}}
\newcommand{\Aq}{\overline{a}_{\rm s}}
\newcommand{\al}{\alpha_{\rm s}}
\newcommand{\Be}{\beta(\Aq)}
\newcommand{\Cr}{C_{\rm R}}
\newcommand{\Ck}{C_{\rm K}}
\newcommand{\Ds}{\Delta_{\rm S}}
\newcommand{\MS}{\overline{\rm MS}}
\newcommand{\Ze}[1]{\zeta_{#1}}
\newcommand{\Df}[2]{\mbox{$\frac{#1}{#2}$}}
\newcommand{\ee}{{\rm e}^+{\rm e}^-}

\noindent{\bf Abstract}
We have discovered 7 intimate connections between the published results for
the radiative corrections, $\Ck$, to the Gross--Llewellyn Smith (GLS) sum
rule, in deep-inelastic lepton scattering, and the radiative corrections,
$\Cr$, to the Adler function of the flavour-singlet vector current, in
$\ee$ annihilation. These include a surprising relation between the
scheme-independent single-electron-loop contributions to the 4-loop QED
$\beta$\/-function and the zero-fermion-loop abelian terms in the 3-loop
GLS sum rule. The combined effect of all 7 relations is to give the
factorization of the 2-loop $\beta$\/-function in
\[\Ds\equiv\Ck\Cr-1=\frac{\Be}{\Aq}\left\{S_1\Cf\Aq+\left[S_2\Tf\Nf
+\Sa\Ca+\Sf\Cf\right]\Cf\Aq^2\right\}+O(\Aq^4)\,,\]
where $\Aq=\al(\mu^2=Q^2)/4\pi$ is the $\MS$ coupling of an arbitrary
colour gauge theory, and
\[S_1=-\Df{21}{2}+12\Ze3\,;\quad S_2=\Df{326}{3}-\Df{304}{3}\Ze3\,;\quad
\Sa=-\Df{629}{2}+\Df{884}{3}\Ze3\,;\quad \Sf=\Df{397}{6}+136\Ze3-240\Ze5\]
specify the sole content of $\Ck$ that is not already encoded in $\Cr$ and
$\Be=Q^2\rd\Aq/\rd Q^2$ at $O(\Aq^3)$. The same result is obtained by
combining the radiative corrections to Bjorken's polarized sum rule with
those for the Adler function of the non-singlet axial current. We suggest
possible origins of $\beta$ in the `Crewther discrepancy', $\Ds$, and
determine $\Ds/(\Be/\Aq)$, to all orders in $\Nf\Aq$, in the large-$\Nf$
limit, obtaining the {\em entire\/} series of coefficients of which $S_1$
and $S_2$ are merely the first two members.
                                            \vfill\begin{flushleft}
INR-820/93 \\
OUT--4102--45 \\

                            \end{flushleft}
                                            \footnoterule\noindent
$^1$) {\em In memoriam\/} Sergei Grigorievich Gorishny, 1958--1988\\
$^2$) D.Broadhurst@open.ac.uk\\
$^3$) Kataev@inucres.msk.su
\newpage
\setcounter{page}{1}

\newcommand{\As}{a_{\rm s}}
\newcommand{\ep}{\varepsilon}
\newcommand{\ri}{{\rm i}}
\newcommand{\re}{{\rm e}}
\newcommand{\Nc}{N_{\rm C}}
\newcommand{\Za}{Z_{\rm A}}
\newcommand{\Cb}{C_{\rm Bjp}}
\newcommand{\Cg}{C_{\rm GLS}}
\newcommand{\Dv}{D^{\rm V}}
\newcommand{\Da}{D^{\rm A}}
\newcommand{\PI}[2]{\Pi^{\rm #1}_{\rm #2}}
\newcommand{\Vt}{\PI{V}{T}}
\newcommand{\Vl}{\PI{V}{L}}
\newcommand{\At}{\PI{A}{T}}
\newcommand{\Al}{\PI{A}{L}}
\newcommand{\ggg}[6]{
\left|\begin{array}{ccc}
g^#1_#4 & g^#1_#5 & g^#1_#6 \\
g^#2_#4 & g^#2_#5 & g^#2_#6 \\
g^#3_#4 & g^#3_#5 & g^#3_#6
\end{array}\right|}
\newcommand{\gggg}[8]{
\left|\begin{array}{cccc}
g^#1_#5 & g^#1_#6 & g^#1_#7 & g^#1_#8\\
g^#2_#5 & g^#2_#6 & g^#2_#7 & g^#2_#8\\
g^#3_#5 & g^#3_#6 & g^#3_#7 & g^#3_#8\\
g^#4_#5 & g^#4_#6 & g^#4_#7 & g^#4_#8
\end{array}\right|}

\section{Introduction}

In 1972, Crewther~\cite{RJC} related three fundamental constants of an
arbitrary parton model: the anomalous constant $S$, associated with the
amplitude for $\pi^0\to\gamma\gamma$ decay~\cite{ABJ}; the coefficient $K$
in Bjorken's sum rule for polarized deep-inelastic electron
scattering~\cite{BJP}; and the constant $R^\prime$ in the annihilation
channel, giving the asymptotic value of the Adler function~\cite{SLA} for
the correlator of the iso-vector axial current. His non-perturbative
derivation relied on conformal and chiral invariance of the leading
short-distance singularity, with coefficient $S$, in the operator product
expansion (OPE) of the 3-point function $A V V$, for $\pi_0$ decay, where
$A=J^{\mu}_5$ is the iso-vector axial current and $V=J^{\mu}_{\rm EM}$ is
the electromagnetic current. To obtain $3S=K R^\prime$~\cite{RJC}, one
first takes the OPE of the 2-point function $V V$, in which one encounters
the axial current, $A$, with the coefficient $K$ of Bjorken's polarized sum
rule. Then one obtains $R^\prime$ in the leading term of the resultant OPE
of the $A A$-correlator, corresponding to the Adler function of the
iso-vector axial current.

The relation $3S=K R^\prime$ is, necessarily, satisfied by the standard
quark-parton model, which gives $S=\frac12$, $K=1$, $R^\prime=\frac32$, for
an iso-doublet of $u$ and $d$ quarks, each having $\Nc=3$ colours. The
chiral symmetry of the quark-parton model means that $K$ also occurs in the
Gross--Llewellyn Smith (GLS) sum rule~\cite{GLS} of deep-inelastic neutrino
scattering, corresponding to the coefficient of the vector current in the
vector-axial correlator. Invoking both chiral symmetry and SU$(3)$-flavour
symmetry~\cite{RJC}, one obtains $R^\prime=\frac34R$, where $R=2$
corresponds to the Adler function of the $V V$-correlator, giving the
quark-parton model prediction for $\ee$ annihilation (below the charm
threshold).

It is not at all clear what the theoretical status and phenomenological
consequences of the Crewther connection might be in QCD, where radiative
corrections to the naive quark-parton model give a dependence on the
running coupling $\Aq=\al(\mu^2=Q^2)/4\pi$ that is appreciable at presently
accessible values of $-q^2\equiv Q^2$. In this paper we study
radiative corrections to deep-inelastic~\cite{LAV} and
annihilation~\cite{GKL} processes, at next-to-next-to-leading order (NNLO)
in the $\MS$ scheme, discovering that they are intimately connected, in a
manner that is profoundly related to the Crewther connection~\cite{RJC}.

\section{Deep-inelastic and annihilation results at NNLO}

In deep-inelastic lepton scattering, radiative corrections to the GLS sum
rule~\cite{GLS},
\begin{equation}
\Df{1}{2}\int_{0}^{1}\rd x\,F_3^{\overline{\nu}{\rm p}+\nu{\rm p}}(x,Q^2)
=3\,\Cg(\Aq)\,,
\label{gls}
\end{equation}
and to Bjorken's polarized sum rule~\cite{BJP},
\begin{equation}
\int_{0}^{1}\rd x\,g_{1}^{{\rm e p}-{\rm e n}}(x,Q^2)
=\frac13\left|\frac{g_{\rm A}}{g_{\rm V}}\right|\Cb(\Aq)\,,
\label{bjp}
\end{equation}
have been obtained~\cite{LAV} to NNLO in the $\MS$ scheme, where the
dependence on $Q^2=-q^2$ is absorbed, at large $Q^2$, into the
coupling $\Aq=\al(\mu^2=Q^2)/4\pi$. A difference between $\Cg$ and
$\Cb$ is first encountered at $O(\Aq^3)$, where so called
`light-by-light-type'
diagrams contribute to $\Cg$, but not to $\Cb$.

In annihilation, at large $Q^2$, the Adler functions $\Dv_{\rm EM}$ and
$\Da_{\rm NS}$ have been calculated~\cite{GKL} at NNLO, again in the $\MS$
scheme, for the electromagnetic current $J^{\mu}_{\rm EM}=\sum_f Q_f
\overline{\psi}_f\gamma^\mu\psi_f$, where the sum is over $\Nf$ flavours of
quark, with charges $Q_f$, and for the {\em non\/}-singlet axial current
$J^\mu_5=\overline{\psi}_i\gamma^\mu\gamma_5\psi_j$, where $i$ and $j$ are
{\em different\/} quark flavours. In~\cite{GKL} it was assumed that
$\Da_{\rm NS}$, relevant to $\tau$ decay, is identical to the Adler
function $\Dv_{\rm NS}$, of the non-singlet vector current, whose radiative
corrections are obtained by omitting the NNLO light-by-light-type terms that
contribute to $\Dv_{\rm EM}$. Analytic continuation of $\Dv_{\rm EM}$ and
$\Da_{\rm NS}$ to the time-like region yields contributions to the processes
$\ee\to{\rm hadrons}$ and $\tau\to\nu_\tau+{\rm hadrons}$. Note that the
correlator of the {\em singlet\/} axial current, related to $Z^0\to{\rm
hadrons}$, receives anomalous contributions from diagrams with
gluons in the intermediate state, considered in~\cite{KAK,CAK}.

To investigate the possibility of a perturbative Crewther connection, in
colour gauge theories, we study the radiative corrections
\begin{eqnarray}
\Ck&\equiv&\Cg(\Aq)=1-3\Cf\Aq+O(\Aq^2)\,,
\label{ck}\\
\Cr&\equiv&\;\frac{\Dv(\Aq)}{\Nf\Nc}\;
=1+3\Cf\Aq+O(\Aq^2)\,,\quad
\Dv\equiv-12\pi^2Q^2\frac{\rd\Pi^{\rm V}}{\rd Q^2}\,,
\label{cr}
\end{eqnarray}
to the GLS sum rule~(\ref{gls}) and the Adler function, $\Dv$, of the
correlator $(q_\mu q_\nu-q^2g_{\mu\nu})\Pi^{\rm V}$
of the flavour-singlet vector current $J_\mu=\sum_f
\overline{\psi}_f\gamma_\mu\psi_f$, with $\Nf$ active quark flavours.

The NNLO results for $\Ck$~\cite{LAV} and $\Cr$~\cite{GKL}, in the $\MS$
scheme, are given in Table~1, for an arbitrary colour gauge group. (The
colour factors take the values $\Tf=\frac12$, $\Ca=\Nc=3$, $\Cf=\frac43$,
$d_{abc}^2=\frac{40}{3}$, in the particular case of QCD.) To obtain the
radiative corrections~(\ref{cr}), to the flavour-singlet Adler function
$\Dv$, one has merely to give the quarks equal charges, $Q_f={}$constant,
in $\Dv_{\rm EM}$, corresponding to setting $(\sum_f Q_f)^2=\Nf\sum_f
Q_f^2$ in the results of~\cite{GKL}. (Note that we consider $\Dv$ at large
space-like $Q^2=-q^2$ and hence omit the $\pi^2$ terms of Eq.~(12)
of~\cite{GKL}, which result from analytic continuation to the time-like
region.) The NNLO radiative corrections to $\Cb$ are obtained by dropping
light-by-light-type terms, proportional to $d_{abc}^2$, from $\Ck\equiv\Cg$;
the corrections to $\Da_{\rm NS}/\Nc=\Dv_{\rm NS}/\Nc$ are obtained by
dropping them from $\Cr\equiv\Dv/\Nf\Nc$.

At first glance, one is tempted to conclude from~\cite{RJC} that the
radiative corrections~(\ref{ck},\ref{cr}) might give a product, $\Ck\Cr$,
that is free of radiative corrections, in the spirit of the
no-renormalization theorem~\cite{ABT} for the axial anomaly that determines
$\pi_0$ decay in the zero-mass limit. However, the NLO corrections to
$\Ck$~\cite{GAL} and $\Cr$~\cite{CKT} give $\Ck\Cr\neq1$, when one absorbs
the $\al^2\ln(Q^2/\mu^2)$ term of each process in the $\MS$ coupling
$\Aq=\al(\mu^2=Q^2)/4\pi$. The recent availability of highly non-trivial
NNLO results, for both $\Ck$~\cite{LAV} and $\Cr$~\cite{GKL}, prompted us
to study the `Crewther discrepancy' $\Ds\equiv(\Ck\Cr-1)$, at $O(\Aq^3)$.

\section{Anatomy of a discovery}

After much investigation of Table~1, we discovered the following remarkable
relation between the $\MS$ results of~\cite{LAV,GKL}, for any colour gauge
theory, renormalized at $\mu^2=Q^2$:
\begin{equation}
\Ds\equiv\Ck\Cr-1=\frac{\Be}{\Aq}\left\{S_1\Cf\Aq+\left[S_2\Tf\Nf
+\Sa\Ca+\Sf\Cf\right]\Cf\Aq^2\right\}+O(\Aq^4)\,,
\label{rmk}
\end{equation}
where $\Be\equiv Q^2\rd\Aq/\rd Q^2=\Aq\sum_{n\geq1}\beta_n\Aq^n$, and
\begin{equation}
S_1=-\Df{21}{2}+12\Ze3\,;\quad S_2=\Df{326}{3}-\Df{304}{3}\Ze3\,;\quad
\Sa=-\Df{629}{2}+\Df{884}{3}\Ze3\,;\quad \Sf=\Df{397}{6}+136\Ze3-240\Ze5
\label{Sis}
\end{equation}
specify the sole NNLO content of $\Ck$ that is not derivable from
$\Cr$ and from the coefficients
\begin{equation}
\beta_1=-\Df{11}{3}\Ca+\Df43\Tf\Nf\,;\quad
\beta_2=-\Df{34}{3}\Ca^2+\Df{20}{3}\Ca\Tf\Nf+4\Cf\Tf\Nf
\label{b12}
\end{equation}
of the two-loop $\beta$\/-function. The {\em same\/} result is obtained by
combining the NNLO corrections to Bjorken's polarized sum rule~(\ref{bjp})
with those for the Adler function $\Da_{\rm NS}$ of the non-singlet axial
current, which differ from $\Ck$ and $\Cr$, respectively, merely by
omitting $d_{abc}^2\Aq^3$ terms that cancel in~(\ref{rmk}).

Since $\Ck$ and $\Cr$, taken up to $O(\Aq^3)$, each involve the 11 distinct
colour factors of the terms $\{T_n|\,n=1,11\}$, defined in Table~1, the
existence of a relation of the form of~(\ref{rmk}) entails the following
`seven wonders' of the Crewther discrepancy $\Ds\equiv(\Ck\Cr-1)$:
\begin{enumerate}
\item The leading-order terms cancel in $\Ds$.
\item The NLO corrections give no $\Cf^2\Aq^2$ term in $\Ds$.
\item The NLO corrections give $\Cf\Ca\Aq^2$ and $\Cf\Tf\Nf\Aq^2$ terms in
$\Ds$ that are in the same ratio as the $\Ca$ and $\Tf\Nf$
terms in $\beta_1$.
\item The NNLO corrections give no $\Cf^3\Aq^3$ term in $\Ds$. This leads
to the astonishing observation that the scheme-independent
single-electron-loop contributions, $\beta^{[1]}_{\rm QED}$, to the QED
$\beta$\/-function, are obtained, up to 4-loops, by taking the {\em
reciprocal\/} of the zero-fermion-loop abelian terms in the 3-loop GLS
result of~\cite{LAV}, giving
\begin{equation}
\beta^{[1]}_{\rm QED}(a)=
\frac{\frac43a^2}{1-3a+\frac{21}{2}a^2-\frac32{a^3}
+O(a^4)}=\Df43a^2+4a^3-2a^4-46a^5+O(a^6)\,,
\label{ast}
\end{equation}
in precise agreement with~\cite{QED}. Such is the power of
relation~(\ref{rmk}).
\item The NNLO light-by-light-type terms of~\cite{LAV} and~\cite{GKL},
involving $T_{11}\equiv\frac{\Nf}{\Nc}d_{abc}^2\Aq^3$, cancel in $\Ds$
(taking equal quark charges in~\cite{GKL}, to obtain the singlet
Adler function $\Dv$).
\item The NNLO corrections in $\Ds$ are expressible as the sum of multiples
of the one-loop and two-loop contributions, $\beta_1\Aq^2$ and
$\beta_2\Aq^3$, to the $\beta$\/-function.
\item At NNLO, $\beta_2\Aq^3$ occurs in $\Ds$ with the {\em same\/}
coefficient that multiplies $\beta_1\Aq^2$,
at NLO, allowing one to factor out $\Be$ in~(\ref{rmk}).
Moreover, this factorization is {\em independent\/} of the momenta in the
two processes: if one takes the GLS sum-rule results at a momentum
transfer $-q^2=Q_{\rm K}^2$, and the Adler function $\Dv$ at $-q^2=Q_{\rm
R}^2$, the factorization of~(\ref{rmk}) still occurs, with the replacements
$\Aq\to\al(\mu^2=Q_{\rm R}^2)/4\pi$ and
\begin{equation}
S_1\to S_1 -3\lambda             \,;\,\,
S_2\to S_2+16\lambda -4\lambda^2 \,;\,\,
\Sa\to \Sa-46\lambda+11\lambda^2 \,;\,\,
\Sf\to \Sf+12\lambda             \,;
\label{rep}
\end{equation}
where $\lambda=\ln(Q_{\rm K}^2/Q_{\rm R}^2)$.
\end{enumerate}

The corresponding 7 relations between the coefficients of $\Ck=\sum_n k_n
T_n+O(\Aq^4)$ and $\Cr=\sum_n r_n T_n+O(\Aq^4)$, given in Table~1, can be
divided into two groups. Observations 1,2,4,5, above, correspond to the 4
conditions
\begin{equation}
0
= k_1+r_1
= k_2+r_2              +k_1r_1
= k_5+r_5              +k_1r_2+r_1k_2
= k_{11}+r_{11}\,,
\label{cz}
\end{equation}
which are required to ensure the absence of terms in $\Ds$ that cannot
occur in the factorization~(\ref{rmk}). Observations 3,6,7, above,
correspond to the 3 remaining conditions
\begin{eqnarray}
  -\Df{3}{11}   (k_3+r_3)
&=&\Df{3}{4}    (k_4+r_4)
\nonumber\\
&=&\Df{1}{7}    (k_7+r_7)
  +\Df{11}{28}  (k_9+r_9)
  +\Df{121}{112}(k_{10}+r_{10})
 \nonumber\\
&=&\Df{1}{11}   (k_6+r_6+k_1r_3+r_1k_3)
  +\Df{1}{4}    (k_8+r_8+k_1r_4+r_1k_4)\,,
\label{cs}
\end{eqnarray}
which relate equivalent ways of evaluating $S_1=-\frac{21}{2}+12\Ze3$,
consistent with~(\ref{rmk},\ref{b12}). We invite any reader who may still
doubt the significance of the factorization of the two-loop
$\beta$\/-function in~(\ref{rmk}) to use the coefficients of Table~1 to
verify that the 7 necessary conditions in~(\ref{cz},\ref{cs}) are
satisfied in a highly non-trivial manner.

We relate our discovery~(\ref{rmk}) to~\cite{RJC} by observing that it
seems rather natural to obtain $\Ck\Cr=1$ at any fixed point, where
$\beta=0$, since Crewther's assumptions of conformal and chiral invariance
should hold in that scale-free limit. We note that the anomalous dimension
of the pseudo-scalar operator $G^{\mu\nu}\widetilde{G}_{\mu\nu}$, in the
anomalous divergence of the singlet axial current, is
$-\beta(\As)/\As$~\cite{SAL}. This makes it reasonable that $\Ck\Cr=1$,
when $\beta=0$, corresponding to {\em no\/} renormalization of the anomaly
at {\em any\/} fixed point, and hence suggesting that one may expect to
find a Crewther discrepancy $\Ds\equiv(\Ck\Cr-1)\propto\beta$.
(See~\cite{SAL,REC} for recent studies of corrections to the one-loop
axial anomaly equation.)

Though we cannot yet be sure that $\beta$ can be factored out of
$(\Ck\Cr-1)$, beyond NNLO, it seems most likely to us that at any given
order, $\Aq^N$, one will encounter in $\Ds$ only the coefficients
$\{\beta_n|\,n<N\}$, multiplied by linear combinations of colour factors.
We believe that it is the scale-dependent procedure of renormalization that
modifies the naive result $\Ds=(\Ck\Cr-1)=0$, suggested by the conformal
arguments of~\cite{RJC} and the essentially one-loop nature of the
anomaly~\cite{ABT}. Since the coefficients $\beta_n$ multiply all
scale-dependent perturbative artefacts of the renormalization procedure,
one expects them to occur in $\Ds$. This does not, of itself, require that
$\Ds\propto\beta$. However, the existence of the relations~(\ref{cs}),
between the highly non-trivial coefficients of Table~1, may be taken as
powerful circumstantial evidence in favour of this stronger hypothesis.

We leave these considerations for later work and now obtain {\em all\/} the
$O(1/\Nf)$ terms of
\begin{equation}
\frac{\Ck\Cr-1}{\Be/\Aq}=\frac{\Cf}{\Tf\Nf}
\sum_{n=1}^\infty S_n(\Tf\Nf\Aq)^n
+O(1/\Nf^2)\,,
\label{slf}
\end{equation}
in the $\MS$ scheme, taking the limit $\Nf\to\infty$, with $\Nf\Aq$
fixed. In obtaining all the coefficients $S_n$, we provide an
all-orders consistency test of the procedures of~\cite{LAV}.

\section{All-orders results at large $\Nf$}

We follow~\cite{LAV,CAK,GAL,SAL,THV,ANT,ZAN} in defining an axial current,
within the framework of dimensional regularization, by calculating Green
functions of the renormalized non-singlet antisymmetric-tensor current
\begin{equation}
A_{\kappa\lambda\mu}\equiv\Df{\ri}{2}\Za
\overline\psi_i(\gamma_\kappa\gamma_\lambda\gamma_\mu
-\gamma_\mu\gamma_\lambda\gamma_\kappa)\psi_j\,,\quad
\Za=1+\sum_{n=1}^{\infty}\As^n\sum_{p=0}^{n-1}\frac{Z_{n,p}}{\ep^p}\,,
\label{Aren}
\end{equation}
where $i$ and $j$ are different quark flavours, $d\equiv4-2\ep$ is the
spacetime dimension, $\As=\al/4\pi$ is the $\MS$ coupling, renormalized at
scale $\mu$, and $\Za$ is a {\em non\/}-minimal renormalization constant,
constructed so as to preserve chiral symmetry and to have {\em vanishing\/}
anomalous dimension. The condition
\begin{equation}
\frac{\rd\ln\Za}{\rd\ln\mu^2}
=\left(-\ep+\frac{\beta(\As)}{\As}\right)
\frac{\rd\ln\Za}{\rd\ln\As}=O(\ep)\
\label{noad}
\end{equation}
thus relates the singular terms in~(\ref{Aren}) to the finite terms, giving
NNLO singular terms
\begin{equation}
Z_{2,1}=\Df12\beta_1Z_{1,0}\,,\quad
Z_{3,2}=\Df13\beta_1^2Z_{1,0}\,,\quad
Z_{3,1}=\Df13\beta_2Z_{1,0}+\Df16\beta_1(Z_{1,0}^2+4Z_{2,0})\,,
\label{zi}
\end{equation}
in terms of $\beta_1$ and $\beta_2$, in~(\ref{b12}), and the NLO finite
terms~\cite{LAV,SAL}
\begin{equation}
Z_{1,0}=-4\Cf\,,\quad Z_{2,0}=22\Cf^2-\Df{107}{9}\Cf\Ca+\Df49\Cf\Tf\Nf\,.
\label{zf}
\end{equation}

The procedure for obtaining Green functions involving the non-singlet axial
current is to combine the non-minimal renormalization of~(\ref{Aren}) with
the standard $\MS$ renormalization of the bare coupling constant, $g_0$.
One then subtracts any polynomial in the momenta that is singular at
$\ep=0$, and takes the limit $\ep\to0$. Thereafter, one multiplies by the
appropriate number of 4-dimensional Levi-Civita tensors (one for each axial
vertex), to obtain renormalized Green functions of the conventional
4-dimensional axial current, which may be written schematically as
\begin{equation}
A_\nu=\overline\psi_i\gamma_\nu\gamma_5\psi_j
\cong\Df{1}{3!}\ep_{\kappa\lambda\mu\nu}\lim_{d\to4}
\left[\Df{\ri}{2}\Za\overline\psi_i
(\gamma^\kappa\gamma^\lambda\gamma^\mu
-\gamma^\mu\gamma^\lambda\gamma^\kappa)\psi_j\right]\,,
\label{sch}
\end{equation}
with the limit $d\to4$ taken {\em after\/} all renormalization. Before
renormalization, all reference to the Levi-Civita tensor, and hence to
$\gamma_5$, is resolutely avoided. This enables covariant $d$\/-dimensional
calculation, albeit at the expense of large traces over
$\gamma$\/-matrices. With traces involving only an even number of
axial vertices, it is presumed~\cite{DKR} that all physical results are
identical to those that would have been obtained by naively using the
4-dimensional identities $\gamma_\mu\gamma_5=-\gamma_5\gamma_\mu$ and
$\gamma_5^2=1$. We now test this in an all-orders calculation.

\subsection{Annihilation processes}

The renormalized correlator of the vector current
$V_\mu\equiv\overline\psi_i\gamma_\mu\psi_j$ may be written as
\begin{equation}
\ri\int\rd x\,\re^{\ri q\cdot x}
\langle T\{V^\mu(x)V^{\dagger}_\nu(0)\}\rangle
=-q^2g^\mu_\nu\Vt+q^\mu q_\nu(\Vl+\Vt)\,,
\label{piv}
\end{equation}
with $\Vl=0$, for massless quarks. The decomposition of the correlator
of~(\ref{Aren}) may be written as
\begin{equation}
\ri\int\rd x\,\re^{\ri q\cdot x}
\langle T\{A^{\alpha\beta\gamma}(x)
A^{\dagger}_{\kappa\lambda\mu}(0)\}\rangle
=-q^2G^{\alpha\beta\gamma}_{\kappa\lambda\mu}\At+
G^{\alpha\beta\gamma\delta}_{\kappa\lambda\mu\nu}q^\nu q_\delta(\Al+\At)\,,
\label{pia}
\end{equation}
with a tensor structure given by the determinants
\begin{equation}
G^{\alpha\beta\gamma}_{\kappa\lambda\mu}
\equiv\ggg{\alpha}{\beta}{\gamma}{\kappa}{\lambda}{\mu}\,,\quad
G^{\alpha\beta\gamma\delta}_{\kappa\lambda\mu\nu}
\equiv\gggg{\alpha}{\beta}{\gamma}{\delta}{\kappa}{\lambda}{\mu}{\nu}\,.
\label{g34}
\end{equation}
For massless quarks, chiral symmetry requires that $\Al$ and $(\At-\Vt)$ be
constants. Note that they may be non-zero, since the non-minimal
renormalization of~(\ref{Aren}), combined with minimal subtraction of
infinities, may still leave chiral-symmetry-breaking finite terms in the
renormalized expressions for divergent quantities. Only the
subtraction-free non-singlet Adler functions are required to be equal:
\begin{equation}
\Da_{\rm NS}=-12\pi^2Q^2\frac{\rd\At}{\rd Q^2}
=-12\pi^2Q^2\frac{\rd\Vt}{\rd Q^2}
=\Dv_{\rm NS}=\Nc(1+3\Cf\Aq+O(\Aq^2))\,.
\label{lead}
\end{equation}
In~\cite{ANT} it was shown that the equality of $\Da_{\rm NS}$ and
$\Dv_{\rm NS}$, at the two-loop level, requires the leading-order
renormalization $\Za=1-4\Cf\As+O(\As^2)$, which then gives an infinite NLO
renormalization in~(\ref{Aren}), with $Z_{2,1}=-2\beta_1\Cf$, according
to~(\ref{zi}). We now investigate the situation, to {\em all\/} orders in
the coupling $\Nf\As$, in the large-$\Nf$ limit.

As $\Nf\to\infty$, with $\Nf\Aq$ fixed, all radiative corrections to the
parton model are suppressed by at least one factor of $1/\Nf$. From the
$O(1/\Nf)$ corrections to $\Dv_{\rm NS}$, we obtain those in
\begin{equation}
\Cr\equiv\frac{\Dv}{\Nf\Nc}=\frac{\Dv_{\rm NS}}{\Nc}+O(1/\Nf^2)
=1+\frac{\Cf}{\Tf\Nf}\sum_{n=1}^\infty R_n(\Tf\Nf\Aq)^n
+O(1/\Nf^2),
\label{rlf}
\end{equation}
in the $\MS$ scheme, with $O(1/\Nf)$ coefficients given in closed form by
\begin{equation}
R_n=\Df32\,4^n(n-1)!\sum_{p=1}^n
\frac{\left(-\Df59\right)^{n-p}}{(n-p)!}\,\frac{\Psi_{p+1}^{[p]}}{(p-1)!}\,,
\label{Rn}
\end{equation}
in terms of the recently obtained momentum-scheme coefficients~\cite{LNF}
\begin{equation}
\Psi_n^{[n-1]}=\frac{(n-1)!}{(-3)^{n-1}}\left[-2n+4-\frac{n+4}{2^n}
+\frac{16}{n-1}\sum_{n/2>s>0}s\left(1-2^{-2s}\right)
\left(1-2^{2s-n}\right)\zeta_{2s+1}\right],
\label{Psi}
\end{equation}
that specify the $O(1/\Nf)$ terms of $\Psi\equiv\beta_{\rm MOM}$, the QED
$\beta$\/-function in the momentum (MOM) subtraction scheme. The all-orders
result~(\ref{Psi}) reproduces the $O(1/\Nf)$ 4-loop results of~\cite{QED},
for $n\leq4$. To transform to the $\MS$ Adler-function
coefficients~(\ref{Rn}), one has merely to observe that, at $O(1/\Nf)$,
a MOM-scheme subtraction at $-q^2=Q^2$ is equivalent to a $\MS$
renormalization at $\mu^2=\re^{-5/3}Q^2$. From~(\ref{Rn},\ref{Psi}), we
readily obtain the results for $R_n$ in Table~2.
The first 3 coefficients agree with~\cite{GKL}; the remainder are new.

\subsection{Axial renormalization constant}

We determine the renormalization constant of~(\ref{Aren}), at $O(1/\Nf)$,
by imposing the chiral-symmetry relations $\Da_{\rm NS}=\Dv_{\rm NS}$ and
$\rd\Al/\rd Q^2=0$. First we calculate the contributions to the axial
correlator of the generic $n$\/-loop bare diagrams with $n-1$ quark
loops, keeping $n$ as an algebraic variable. This result involves an
$F_{3,2}$ hypergeometric function, whose expansion about $\ep=0$ cannot be
effected in terms of $\zeta$\/-functions. Fortunately, coupling-constant
renormalization ensures that one needs the function only in the
limit~\cite{LNF} $\ep\to0$, with $n\ep$ fixed, where it is a tri-gamma
function whose expansion yields $\zeta$\/-functions~\cite{DJB}. Analyzing
the residual, analytically simpler, bare contributions, we encounter the 3
anticipated problems that must be solved by the non-minimal
renormalization~(\ref{Aren}):  there are non-subtractable singular bare
terms, involving $\ln(Q^2/\mu^2)/\ep$; the bare transverse axial and vector
contributions differ by logarithmic terms; the bare longitudinal axial
contributions also have a logarithmic $Q^2$-dependence.

For these problems, a single cure is available: the
renormalization~(\ref{Aren}), which acts only on the one-loop term, at
$O(1/\Nf)$. At $n+1$ loops, only one new constant is at our disposal: the
leading term in the large-$\Nf$ expansion of the coefficient $Z_{n,0}$.
Following the methods of~\cite{LNF}, we have explicitly verified that this
suffices to solve all 3 problems. The required all-orders
solution to~(\ref{noad}) is
\begin{equation}
\Za=1+\frac{\Cf\ep}{6\Tf\Nf}\widehat{{\rm L}}\left\{
\frac{\ln(1-\frac43\Tf\Nf\As/\ep)}{B(2-\ep,2-\ep)B(3-\ep,1+\ep)}\right\}
+O(1/\Nf^2)\,,
\label{eul}
\end{equation}
where $\widehat{{\rm L}}$ is the Laurent operator, which removes
non-singular terms from the perturbative expansion of the term in braces,
in accordance with the Ansatz of~(\ref{Aren}). Note that the Euler
$B$-functions, in~(\ref{eul}), result from the residue at $n=1$ of the
analytical expression for the bare $n$-loop contribution, in much the same
way that the $O(1/\Nf)$ QED $\beta$\/-function of any MS-like scheme
results from a residue at $n=0$~\cite{LNF,PMP}. Expanding~(\ref{eul}) to
order $\As^3$, we verify the large-$\Nf$ terms in the axial-current
renormalization used in~\cite{LAV}. Using it to all orders, we verify the
chiral-symmetry relations $\Da_{\rm NS}=\Dv_{\rm NS}$ and $\rd\Al/\rd
Q^2=0$, at $O(1/\Nf)$ in the $\MS$ scheme.

\subsection{Deep-inelastic processes}

We now calculate all the $O(1/\Nf)$ radiative corrections to the sum
rules~(\ref{gls},\ref{bjp}), which differ only by terms of $O(1/\Nf^2)$.
For the polarized deep-inelastic electron scattering sum rule~(\ref{bjp}),
we calculate the generic $n$\/-loop $O(1/\Nf)$ bare diagrams for forward
Compton scattering of a vector current, with momentum $q$, off a
zero-momentum quark~\cite{LAV,GAL,GLT}, by inserting $n-1$ quark loops into
the one-loop diagrams, obtaining simple $\Gamma$\/-functions. The Compton
diagrams must then be divided~\cite{LAV} by~(\ref{eul}), to obtain the
coefficient $\Cb$ of the axial current~(\ref{Aren}) in the OPE of $V V$.
After the coupling-constant renormalization
\begin{equation}
\left(\frac{g_0}{4\pi}\right)^2=
\left(\frac{\mu^2\re^\gamma}{4\pi}\right)^\ep
\frac{\As}{1-\frac43\Tf\Nf\As/\ep}+O(1/\Nf^2)\,,
\label{lgb}
\end{equation}
we obtain the $O(1/\Nf)$ contributions to
\begin{equation}
\Ck\equiv\Cg=\Cb+O(1/\Nf^2)=
1+\frac{\Cf}{\Tf\Nf}\sum_{n=1}^\infty K_n(\Tf\Nf\Aq)^n
+O(1/\Nf^2)\,,
\label{klf}
\end{equation}
in the $\MS$ scheme, with $O(1/\Nf)$ coefficients given in closed form by
\begin{equation}
K_n=\lim_{z\to0}\left(-\frac43\frac{\rd}{\rd
z}\right)^{n-1}\overline{K}(z)\,,\quad
\overline{K}(z)=-\frac{(3+z)\exp(5z/3)}{(1-z^2)(1-z^2/4)}\,,
\label{Kn}
\end{equation}
where $\overline{K}(z)$ is obtained from the $\ep\to0$ limit of the bare
$n$\/-loop contribution, with $z=n\ep$ fixed. The renormalization
constant~(\ref{eul}) precisely cancels the infinities of the bare terms,
obtained from the residue of the pole at $n=0$. The same results are
obtained for $\Cg$, in the large-$\Nf$ limit, since the diagrams that
distinguish the sum rules have three (or more) gluons in the
$t$\/-channel~\cite{LAV} and hence are (at least) of order $1/\Nf^2$.

{}From~(\ref{Kn}), we readily obtain the results for $K_n$ in Table~2. The
first 3 coefficients agree with~\cite{LAV}; the remainder are new.
Combining~(\ref{Rn},\ref{Kn}), we obtain the $O(1/\Nf)$ $\MS$ coefficients
$S_n=\frac34(K_{n+1}+R_{n+1})$ in~(\ref{slf}), also given in
Table~2, which may be extended, {\em ad libitum}. The first 2 coefficients,
$S_1$ and $S_2$, agree with~(\ref{Sis}); the remainder are new.

\section{Conclusions}

We have discovered the 7 intricate connections of~(\ref{cz},\ref{cs}),
between the highly non-trivial NNLO radiative corrections, $\Ck$ and $\Cr$,
to the GLS sum rule~\cite{LAV} and the Adler function~\cite{GKL} of the
flavour-singlet vector current, given in Table~1. Forming
$\Ds\equiv(\Ck\Cr-1)$, we find the remarkable result of~(\ref{rmk}),
namely a linear function of the coupling, multiplying the two-loop
$\beta$\/-function, and hence reducing from 11 to 4 the number of
independent colour structures in $\Ds$, up to $O(\Aq^3)$. Two of the 4
coefficients in~(\ref{rmk}), namely $S_1$ and $S_2$, have been obtained
{\em ab initio\/}, as the first two members of the series of large-$\Nf$
coefficients in~(\ref{slf}), whose higher-order members can be obtained
from our new all-orders results~(\ref{Rn},\ref{Kn}). Values are given in
Table~2 for $n<10$.

The consistency of the prescription~(\ref{sch}) with chiral symmetry has
been demonstrated, in the large-$\Nf$ limit, using the all-orders axial
renormalization constant~(\ref{eul}). A validation of $\Sa$ and $\Sf$, the
$O(1/\Nf^2)$ coefficients in~(\ref{rmk}), has not been attempted here. We
note, however, that recent progress with $O(1/\Nf^2)$ corrections to
QED~\cite{JAG}, and to the Gross-Neveu model~\cite{GNV}, suggests that one
may eventually be able to obtain the entire series of coefficients that
have $\Sa$ and $\Sf$ as their leading members. In any case, the
relations~(\ref{cz},\ref{cs}) give one a high degree of confidence in the
accuracy of~\cite{LAV,GKL}. Section~3 offers some general observations,
suggesting that the relation $\Ds\propto\beta$ may persist beyond NNLO. In
any case, we confidently expect the expansion of $\Ds$ to involve only the
coefficients of the $\beta$\/-function, multiplied by linear combinations
of colour factors.

We believe that the hypothesis $\Ds\propto\beta$ merits
close attention. Its proof (or disproof) could contribute significantly to
an area of field theory that combines deep principles~\cite{RJC,ABT} with
calculational achievement~\cite{LAV,GKL,ZAN,BKT} and phenomenological
relevance~\cite{CKL}. We recommend careful re-examination of the Crewther
connection~\cite{RJC} in the light of our findings.
The new connections~(\ref{rmk},\ref{ast}) suggest that
gauge theories know much more about it than has been supposed.

\noindent{\bf Acknowledgments}
We thank Bob Jaffe for asking us about the status of~\cite{RJC} in QCD. ALK
thanks Peter White for help in organizing his recent fruitful tour in
England, contributing to the appearance of this work, which is dedicated
to the memory of his late friend and colleague S.\ G.\ Gorishny,
who mentioned to him, in private discussions in 1987, not long before
his death from cancer, that
``calculations of the NNLO corrections to the GLS sum rule may be connected
to problems of the manifestation of the anomaly and of the careful
treatment of its renormalization within dimensional regularization''.
\newpage
\raggedright
\newcommand{\lb}{$\\[3pt]$\quad{}}
\newcommand{\od}[2]{+\Bigl[#2\Bigr]x^{#1}}
\newcommand{\ox}[2]{#2x^{#1}}
\newcommand{\hr}{\vskip0.3cm\hrule\vskip0.2cm}
\newcommand{\Pm}{\phantom{-}}
{\bf Table 1}~~NNLO results of~\cite{LAV,GKL} for
$\Ck=\sum_n k_n T_n +O(\Aq^4)$,
$\Cr=\sum_n r_n T_n +O(\Aq^4)$.
\[\begin{array}{rrll}n& T_n & \Pm k_n & \Pm r_n                  \\\hline
 1 &          \Cf\Aq  & -3
                      & \Pm3                                     \\\hline
 2 &        \Cf^2\Aq^2& \Pm\frac{21}2
                      & -\frac32                                 \\
 3 &       \Cf\Ca\Aq^2& -23
                      & \Pm\frac{123}{2}-44\Ze3                  \\
 4 &    \Cf\Tf\Nf\Aq^2& \Pm8
                      & -22+16\Ze3                               \\\hline
 5 &        \Cf^3\Aq^3& -\frac32
                      & -\frac{69}{2}                            \\
 6 &     \Cf^2\Ca\Aq^3& \Pm\frac{1241}{9}-\frac{176}{3}\Ze3
                      & -127-572\Ze3+880\Ze5                     \\
 7 &     \Cf\Ca^2\Aq^3& -\frac{10874}{27}+\frac{440}{3}\Ze5
                      & \Pm\frac{90445}{54}-\frac{10948}{9}\Ze3
                        -\frac{440}{3}\Ze5                       \\
 8 &  \Cf^2\Tf\Nf\Aq^3& -\frac{133}{9}-\frac{80}{3}\Ze3
                      & -29+304\Ze3-320\Ze5                      \\
 9 & \Cf\Ca\Tf\Nf\Aq^3& \Pm\frac{7070}{27}+48\Ze3
                        -\frac{160}{3}\Ze5
                      & -\frac{31040}{27}+\frac{7168}{9}\Ze3
                        +\frac{160}{3}\Ze5                       \\
10 &\Cf\Tf^2\Nf^2\Aq^3& -\frac{920}{27}
                      & \Pm\frac{4832}{27}-\frac{1216}{9}\Ze3    \\
11 &\frac{\Nf}{\Nc}
        d_{abc}^2\Aq^3& -\frac{11}{3}+8\Ze3
                      & \Pm\frac{11}{3}-8\Ze3                    \\\hline
\end{array}\]
\vskip0.5cm
{\bf Table 2}~~Large-$\Nf$ expansions~(\ref{rlf},\ref{klf},\ref{slf}),
obtained from~(\ref{Rn},\ref{Kn}), with $x\equiv\Tf\Nf\Aq$.
\hr
$\sum_{n<10}\;R_n x^n=
3x
\od2{
-{22}
+{16}\Ze3}
\od3{
 {4832\over27}
-{1216\over9}\Ze3}
\od4{
-{392384\over243}
+{25984\over27}\Ze3
+{1280\over3}\Ze5}\lb
\od5{
 {11758720\over729}
-{5073920\over729}\Ze3
-{194560\over27}\Ze5}
\od6{
-{3499697920\over19683}
+{357201920\over6561}\Ze3
+{20787200\over243}\Ze5
+{71680\over3}\Ze7}\lb
\od7{
 {381559797760\over177147}
-{9308446720\over19683}\Ze3
-{2029568000\over2187}\Ze5
-{5447680\over9}\Ze7}\lb
\od8{
-{5056220794880\over177147}
+{2445582254080\over531441}\Ze3
+{200033075200\over19683}\Ze5
+{814858240\over81}\Ze7
+{194969600\over81}\Ze9}\lb
\od9{
 {5908327309475840\over14348907}
-{239732713062400\over4782969}\Ze3
-{20850920652800\over177147}\Ze5
-{318236262400\over2187}\Ze7
-{59270758400\over729}\Ze9}$
\hr
$\sum_{n<10}\;K_n x^n=
-3x
\ox2{
+{8}}
\ox3{
-{920\over27}}
\ox4{
+{38720\over243}}
\ox5{
-{238976\over243}}
\ox6{
+{130862080\over19683}}
\ox7{
-{10038092800\over177147}}\lb
\ox8{
+{274593587200\over531441}}
\ox9{
-{82519099473920\over14348907}}$
\hr
$\sum_{n<10}\;S_n x^n=
\Bigl[-{21\over2}
+{12}\Ze3\Bigr]x
\od2{
 {326\over3}
-{304\over3}\Ze3}
\od3{
-{9824\over9}
+{6496\over9}\Ze3
+{320}\Ze5}\lb
\od4{
 {2760448\over243}
-{1268480\over243}\Ze3
-{48640\over9}\Ze5}
\od5{
-{280736320\over2187}
+{89300480\over2187}\Ze3
+{5196800\over81}\Ze5
+{17920}\Ze7}\lb
\od6{
 {10320047360\over6561}
-{2327111680\over6561}\Ze3
-{507392000\over729}\Ze5
-{1361920\over3}\Ze7}\lb
\od7{
-{3723517199360\over177147}
+{611395563520\over177147}\Ze3
+{50008268800\over6561}\Ze5
+{203714560\over27}\Ze7
+{48742400\over27}\Ze9}\lb
\od8{
 {485484017500160\over1594323}
-{59933178265600\over1594323}\Ze3
-{5212730163200\over59049}\Ze5
-{79559065600\over729}\Ze7
-{14817689600\over243}\Ze9}\lb
\od9{
-{7616109282344960\over1594323}
+{726735764193280\over1594323}\Ze3
+{195646580326400\over177147}\Ze5
+{1120185221120\over729}\Ze7\lb
+{316630630400\over243}\Ze9
+{7821721600\over27}\Ze{11}}$
\hr


\newpage

\begin{thebibliography}{99}

\bibitem{RJC}
R.\ J.\ Crewther,
{\sl Phys.\ Rev.\ Lett.}\ {\bf 28} (1972) 1421.

\bibitem{ABJ}
S.\ L.\ Adler,
{\sl Phys.\ Rev.}\ {\bf 177} (1969) 2426;\\
J.\ S.\ Bell and R.\ Jackiw,
{\sl Nuovo Cimento}\ {\bf 60A} (1969) 47.

\bibitem{BJP}
J.\ D.\ Bjorken,
{\sl Phys.\ Rev.}\ {\bf 148} (1966) 1467;
{\bf D1} (1970) 1376.

\bibitem{SLA}
S.\ L.\ Adler,
{\sl Phys.\ Rev.}\ {\bf D10} (1974) 3714.

\bibitem{GLS}
D.\ J.\ Gross and C.\ H.\ Llewellyn Smith,
{\sl Nucl.\ Phys.}\ {\bf B14} (1969) 337.

\bibitem{LAV}
S.\ A.\ Larin and J.\ A.\ M.\ Vermaseren,
{\sl Phys.\ Lett.}\ {\bf B259} (1991) 345.

\bibitem{GKL}
S.\ G.\ Gorishny, A.\ L.\ Kataev and S.\ A.\ Larin,
{\sl Phys.\ Lett.}\ {\bf B259} (1991) 144;\\
{\sl Pisma ZhETF} {\bf 53} (1991) 121.

\bibitem{KAK}
B.\ A.\ Kniehl and J.\ H.\ K\"{u}hn,
{\sl Nucl.\ Phys.}\ {\bf B329} (1990) 547.

\bibitem{CAK}
K.\ G.\ Chetyrkin and A.\ Kwiatkowski,
{\sl Phys.\ Lett.}\ {\bf B305} (1993) 285.

\bibitem{ABT}
S.\ L.\ Adler and W.\ Bardeen,
{\sl Phys.\ Rev.}\ {\bf 182} (1969) 1517.

\bibitem{GAL}
S.\ G.\ Gorishny and S.\ A.\ Larin,
{\sl Phys.\ Lett.}\ {\bf B172} (1986) 109.

\bibitem{CKT}
K.\ G.\ Chetyrkin, A.\ L.\ Kataev and F.\ V.\ Tkachov,
{\sl Phys.\ Lett.}\ {\bf B85} (1979) 277;\\
M.\ Dine and J.\ Sapirstein,
{\sl Phys.\ Rev.\ Lett.}\ {\bf 43} (1979) 668;\\
W.\ Celmaster and R.\ Gonsalves,
{\sl Phys.\ Rev.\ Lett.}\ {\bf 44} (1980) 560.

\bibitem{QED}
S.\ G.\ Gorishny, A.\ L.\ Kataev, S.\ A.\ Larin and L.\ R.\ Surguladze,\\
{\sl Phys.\ Lett.}\ {\bf B256} (1991) 81.

\bibitem{SAL}
S.\ A.\ Larin,
{\sl Phys.\ Lett.}\ {\bf B303} (1993) 113.

\bibitem{REC}
R.\ Akhoury and S.\ Titard,
Univ.\ Michigan preprint UM--TH--91--21 (1991);\\
G.\ T.\ Gabadadze and A.\ A.\ Pivovarov,
{\sl Pisma ZhETF} {\bf 54}  (1991) 305;\\
{\sl Yad. Fiz.}\ {\bf 56} (1993) 257;\\
M.\ Bos, UCLA preprint UCLA--92--TEP--41 (1992).

\bibitem{THV}
G.\ 't Hooft and M.\ Veltman,
{\sl Nucl.\ Phys.}\ {\bf B44} (1972) 189;\\
D.\ A.\ Akyeampong and R.\ Delburgo,
{\sl Nuovo Cimento} {\bf 17A} (1973) 578;\\
P.\ Breitenlohner and D.\ Maison,
{\sl Comm.\ Math.\ Phys.}\ {\bf 52} (1977) 11.

\bibitem{ANT}
I.\ Antoniadis,
{\sl Phys.\ Lett.}\ {\bf B84} (1979) 223;\\
T.\ L.\ Trueman,
{\sl Phys.\ Lett.}\ {\bf B88} (1979) 331.

\bibitem{ZAN}
E.\ B.\ Zijlstra and W.\ L.\ van Neerven,
{\sl Phys.\ Lett.}\ {\bf B297} (1992) 377.

\bibitem{DKR}
D.\ Kreimer,
{\sl Phys.\ Lett.}\ {\bf B237} (1990) 59;\\
P.\ A.\ Baikov and V.\ A.\ Ilyin,
{\sl Teor.\ Mat.\ Fiz.}\ {\bf 88} (1991) 163.

\bibitem{LNF}
D.\ J.\ Broadhurst,
{\sl Z.\ Phys.}\ {\bf C58} (1993) 339.

\bibitem{DJB}
D.\ J.\ Broadhurst,
{\sl Z.\ Phys.}\ {\bf C32} (1986) 249;\\
D.\ T.\ Barfoot and D.\ J.\ Broadhurst,
{\sl Z.\ Phys.}\ {\bf C41} (1988) 81.

\bibitem{PMP}
A.\ Palanques--Mestre and P.\ Pascual,
{\sl Comm.\ Math.\ Phys.}\ {\bf 95} (1984) 277.

\bibitem{GLT}
S.\ G.\ Gorishny, S.\ A.\ Larin and F.\ V.\ Tkachov,
{\sl Phys.\ Lett.}\ {\bf B124} (1983) 217;\\
S.\ G.\ Gorishny and S.\ A.\ Larin,
{\sl Nucl.\ Phys.}\ {\bf B283} (1987) 452.

\bibitem{JAG}
J.\ A.\ Gracey,
{\sl Mod.\ Phys.\ Lett.}\ {\bf A7} (1992) 1945.

\bibitem{GNV}
J.\ A.\ Gracey,
Liverpool preprint LTH--312 (1993);\\
S.\ E.\ Derkachov, N.\ A.\ Kivel, A.\ S.\ Stepanenko and A.\ N.\ Vasil'ev,\\
Saclay preprint SPHT--93--016 (1993).

\bibitem{BKT}
D.\ J.\ Broadhurst, A.\ L.\ Kataev and O.\ V.\ Tarasov,
{\sl Phys.\ Lett.}\ {\bf B298} (1993) 445;\\
T.\ Kinoshita,
{\sl Phys.\ Rev.}\ {\bf D47} (1993) 5013

\bibitem{CKL}
J.\ Ch\'{y}la, A.\ L.\ Kataev and S.\ A.\ Larin,
{\sl Phys.\ Lett.}\ {\bf B267} (1991) 269;\\
J.\ Ch\'{y}la and A.\ L.\ Kataev,
{\sl Phys.\ Lett.}\ {\bf B297} (1992) 385;\\
E.\ Braaten, S.\ Narison and A.\ Pich,
{\sl Nucl.\ Phys.}\ {\bf B373} (1992) 581.


\end{thebibliography}
\end{document}